\documentclass[manuscript]{emulateapj}
\slugcomment{accepted to ApJ Letters}

\shorttitle{Dust Coagulation and Settling in Layered Protoplanetary Disks}
\shortauthors{Ciesla}

\begin{document}


\title{Dust Coagulation and Settling in Layered Protoplanetary Disks}


\author{Fred J. Ciesla}
\affil{Department of Terrestrial Magnetism, Carnegie Institution of Washington,
5241 Broad Branch Road NW, Washington, DC 20015}
\email{ciesla@dtm.ciw.edu}
%\author{C. D. Biemesderfer\altaffilmark{4,5}}
%\affil{National Optical Astronomy Observatories, Tucson, AZ 85719}
%\email{aastex-help@aas.org}

%\and

%\author{R. J. Hanisch\altaffilmark{5}}
%\affil{Space Telescope Science Institute, Baltimore, MD 21218}


%\altaffiltext{1}{Visiting Astronomer, Cerro Tololo Inter-American Observatory.
%CTIO is operated by AURA, Inc.\ under contract to the National Science
%\altaffiltext{2}{Society of Fellows, Harvard University.}
%\altaffiltext{3}{present address: Center for Astrophysics,
%    60 Garden Street, Cambridge, MA 02138}
%\altaffiltext{4}{Visiting Programmer, Space Telescope Science Institute}
%\altaffiltext{5}{Patron, Alonso's Bar and Grill}


\begin{abstract}
Previous models of dust growth in protoplanetary disks considered either uniformly 
laminar or turbulent disks.  This Letter explores how dust growth occurs in a 
layered protoplanetary disk in which the magnetorotational instability generates
turbulence only in the surface layers of a disk.  Two cases are considered: 
a completely laminar dead zone and a dead zone in which turbulence is ``stirred up''
from the MRI acting above.  It is found that dust is depleted from high altitudes
in layered disks faster than in those cases of a uniformly laminar or turblent
disks.  This is a result of the accelerated growth of particles in the turbulent
regions and their storage in the lower levels where they escape energetic
collisions which would result in disruption.  Thus the regions of a protoplanetary disk above
a dead zone would become rapidly depleted in small dust grains, whereas the outer
regions, where the MRI is active troughout, will maintain a small dust 
poplulation at all heights due to the disruptive
collisions and vertical mixing from turbulence.  This structure is similar to 
that which has been inferred for disks around TW Hydra, GM Auriga, and CoKu Tau/4,
which are depleted in dust close to the star, but are optically thick at larger
heliocentric distances.
\end{abstract}


\keywords{turbulence, solar system:formation, planetary systems: protoplanetary disks, stars:pre-main sequence}



\section{Introduction}

The initial stages of planetary system formation involve the coagulation of
dust in protoplanetary disks into planetesimals.  Models of this process have
been developed in order to interpret observations of protoplanetary disks
\citep{dull05,tanaka05,dalessio06} or to understand the details of how this
process occurred in our own solar nebula \citep{weidenschilling80,weidenschilling84,
weidenschilling97}.  An important factor in determining how growth takes place
is whether turbulence is present in the disk, as it will increase the relative
velocities between dust particles, leading to more frequent and energetic 
collisions.  Previous efforts have focused on disks which were uniformly
laminar or uniformly turbulent.  However, the
likely source of turbulence in protoplanetary disks, the magnetorotational
instability (MRI), is expected to only operate in regions of the 
disk which are sufficiently ionized for the gas to couple to the local magnetic 
field \citep{bh91}.  Such regions are expected where the gas temperature 
exceeds $\sim$1000 K allowing the gas to become collisionally ionized or at
the surface layers of the disk where absorption of X-rays and cosmic rays lead to ionization
\citep{gammie96}.  The region which is not ionized sufficiently, which would 
likely be the disk midplane in the planet formation region, is termed the ``dead zone.''

In addition to the energetic radiation flux, the presence
of dust will also determine the ionization fraction of the gas.  \citet{sano00}
demonstrated that grain surfaces allow for rapid recombination of ions and
electrons, thus keeping the ionization fraction low.  Significant ionization 
occurs only after the available surface area of grains is reduced, either through
incorporation of grains into larger bodies, or by the depletion of solids by
settling towards the midplane.

In this Letter, particle growth and settling in a layered
protoplanetary disk is modeled. 
Two cases are considered: that of a laminar
dead zone ($\alpha$=0), and that of a dead zone that becomes turbulent, though to
a lesser extent ($\alpha_{DZ}$=0.1$\alpha_{MRI}$),
in agreement with the models of \citet{fleming03}. The focus here is to demonstrate the effect
that the layered structure has on the vertical distribution of solids
in protoplanetary disk.  A more detailed model,
which explores a wider parameter space will be presented in a future paper.
In \S 2, the model and assumptions are described.  In \S 3, the results of 
different simulations are presented.
The implications for protoplanetary disk observations and structure are discussed in \S4.

\section {Model Description}

The evolution of the number density, $n$, of particles of a 
given mass, $m$, and radius, $a$, at a given height, $z$, above the midplane of 
protoplanetary disk is given by:
\begin{eqnarray}
\frac{\partial n_{m}}{\partial t} & = &
\frac{\partial}{\partial z}
\left( {\cal D}_{z} \rho_{g}\frac{\partial}{\partial z}
\left( \frac{n_{m}}{\rho_{g}} \right)\right)
-\frac{\partial}{\partial z} \left( n_{m} v_{z} \right) \nonumber \\
& & +\frac{1}{2}\int_{0}^{m} K_{m',m-m'}n_{m'}n_{m-m'} dm' \nonumber \\
& & -n_{m}\int_{0}^{\infty} K_{m,m'}n_{m'} dm' + S_{m}
\end{eqnarray}
The first term on the right hand side of the equation describes the vertical diffusion
that takes place due to turbulence, where $D_{z}$ is the diffusivity (taken to be the
local turbulent viscosity of the particles, $\alpha c H$/(1+$St$), where $St$ is the 
local Stokes number of the particle),
and $\rho_{g}$ is the local gas mass density.  The second term on the right hand
side describes the vertical settling due to gravity, where $v_{z}$ is the settling
rate determined by the balance of gravity and the drag force due to 
movement through the gas.  The drag force on the particles is found using the laws given
in \citet{tanaka05}.  Finally, the integral terms on the right hand side of
the equation represent the changes that occur due to coagulation, where the first
term describes the creation of particles of mass $m$ through collisions of smaller
ones, and the second term represents the loss of particles due to incorporation
into larger solids.  The $K$ factors are the kernels, representing
the product of the collisional cross section of the particles, $\pi$($a_{1}$+$a_{2}$)$^{2}$,
with the relative velocities between those particles.  The relative velocities arise
due to Brownian motion, differential settling, gas drag (radial and azimuthal
components), and turbulence (when present).  In
the case of turbulence, the relative velocities are found as described in \citet{dull05}.

Collisions may also result in disruption when velocities are large.  
A simple model is used where collisions are assumed to be destructive
if the total kinetic energy is greater than the strength of the sum of the two
masses, ($m_{1}$+$m_{2}$)$q^{*}$.  In the case of a destructive collision, all of the
mass of the two colliding bodies is assumed to be dispersed into the smallest
sized particles considered here (this is represented by the source term, $S_{m}$,
above).  Here, a
strength of $q^{*}$=$1.25 \times 10^{5}$ erg/g was assumed for all bodies, independent
of size.  This strength corresponds to bodies of equal sizes being disrupted at
collisional velocities in excess of 10 m/s.  This is likely a conservative estimate
as \citet{wurm05} found aggregation can occur at velocities as high as 25 m/s.

The disk considered here is assumed to have surface density and temperature
structures given by $\Sigma$($r$)=2000(1/$r$ AU) g/cm$^{2}$ 
and $T$($r$)=300(1/$r$ AU)$^{0.5}$ K respectively. The disk is assumed to be isothermal 
with height above the disk midplane.  
Inside of the snow line (taken to be 4 AU, where $T$=150 K), dust is 
initially suspended at a solid-to-gas mass ratio of 0.005 with a particle
density of 1.5 g/cm$^{3}$.  All solids are initially distributed as particles 
0.1 $\mu$m in diameter.  In all cases, the disk is assumed to extend to 3 
pressure scale heights, $H$, above the disk midplane.

A simplified model for the growth of the MRI-active region is used here.  The MRI is assumed
to initially be active in the upper 1 g/cm$^{2}$ of a given location of the disk.  As the
available
surface area of the grains decreases, either through coagulation or settling, the active layer
grows.  The MRI is assumed to descend into lower altitudes 
when the total mass denity of dust smaller than $a$=1 $\mu$m in a given cell is less than
10$^{-4}$ times the original mass density of solids \citep{sano00}.  The active
layer is not allowed to expand beyond 100 g/cm$^{2}$, marking the maximum
penetration depth of ionizing radiation, in agreement with 
\citet{gammie96}.  The MRI region is assumed to have a turbulence
characterized by $\alpha$=10$^{-2}$.  In those cases specified, the dead zone develops
a value of $\alpha$=10$^{-3}$ when $\Sigma_{MRI}$/$\Sigma_{DZ}$=0.1 to mimic the
results of \citet{fleming03}.

\begin{figure}[h]
\includegraphics[width=3.1in]{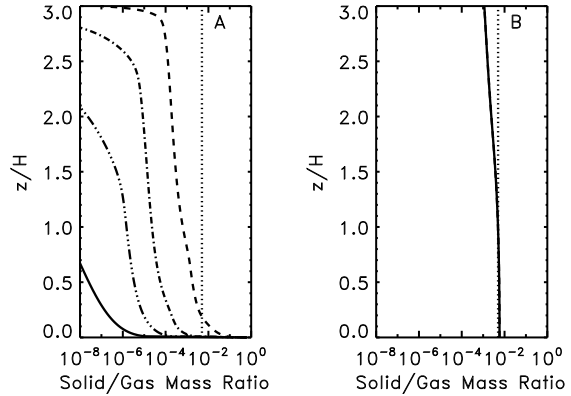}
\caption{Plotted are the vertical distribution of solids at 1 AU for a uniformly 
laminar disk (A) and a uniformly turbulent disk (B) at t=1 (dotted), 1000 (dashed)
10$^{4}$ (dash-dotted), 10$^{5}$ (dash-dot-dot-dotted), and 10$^{6}$ (solid) years.
In the case of the turbulent disk, a steady-state is achieved after $\sim$1000 
years, in which particles are created as readily as they are destroyed.}
\end{figure}

\section{Results}

Figure 1 shows the evolution of the vertical distribution of solids at 1 AU for
the cases of a uniformly laminar disk and a uniformly turbulent disk, which are
the more typical cases considered when modeling dust coagulation in
protoplanetary disks.  In the case of the laminar disk, dust growth originally
occurs due to the brownian motion that the small grains experience in the gas.
As larger grains form, they begin to settle towards the midplane, with higher
settling velocities at higher altitudes.  These larger bodies ``rain out" from 
these high altitudes, sweeping up
the small grains that are present on the way to the midplane.  As a result,
the upper layers of the disk are first depleted in dust, with the lower altitudes
becoming more depleted with time.  Essentially all of the mass of the solids becomes
incorporated into planetesimal-sized bodies which reside at the midplane.

In the uniformly turbulent case ($\alpha$=10$^{-2}$), 
little depletion actually occurs, as turbulence
inhibits the growth of bodies beyond a certain point.  The additional relative
velocity between particles as a result of turublence produces a situation that
has been referred to as the ``meter-sized barrier" \citep{weidenschilling84,
cuzzweid06}.  Turbulence allows particles near 1 meter in size to develop relative
velocities with respect to one another that result in disruptive collisions.
The disruptive timescale for these bodies is less than their growth timescale 
through collisions with smaller particles.  As a result, growth beyond this
size is frustrated, and small dust is constantly resupplied to the nebula
through these collisions and mixed vertically by turbulent diffusion.  Thus,
the vertical distribution of material reaches a steady-state situation after
$\sim$1000 years, where dust is somewhat depleted at high altitudes, but only 
by a factor of a few.

Figure 2 presents the results for layered protoplanetary disks, 
where the turbulent region begins in the extreme upper layers of the disk and 
grows with time as dust is depleted, with both a perfectly laminar dead zone 
and a dead zone which becomes turbulent when the \citet{fleming03} criteria is 
satisfied.  Initially, the dust distributions in these two cases evolve 
identically.  In the MRI-active layers, particle growth is rapid compared to
the dead zone, as turbulence increases the relative velocities between
particles, leading to more frequent collisions.  While growth is somewhat 
frustrated as larger bodies form and collisions become more energetic and 
disruptive, some fraction of larger bodies continuously settle to the deadzone, 
and thus are removed from the MRI-active layer.  As there is no turbulence in 
the dead zone during these early times, and thus no vertical diffusion, there 
is no way to resupply the MRI-active region with solids.  

Particles in the dead zone continue to grow as described for the uniformly
laminar disk, with much of the mass cocentrating at the midplane and being
incorporated into larger bodies.  After 14594 years, the \citet{fleming03} criteria,
where $\Sigma_{MRI}$/$\Sigma_{DZ}$=0.1, is met and turbulence is stirred up in
the dead zone in Case B.  At this time, much of the mass of the solids
are found at the midplane in bodies 5-200 meters in radius, as seen in Figure 3.
This size range is beyond the ``meter-sized" barrier.  Thus, when the
dead zone becomes turbulent, the small dust particles that remain can grow through
mutual collisions, but only to the barrier size, at which point their collisions
become disruptive and maintain a population of fine dust that gets mixed vertically.
However, the turbulence also increases the rate at which dust particles collide
with the larger bodies at the midplane, which results in faster accretion of 
these particles by the larger bodies that have already formed.  As a result,
the growth of the large bodies is then accelerated
by the presence of turbulence since collisions are not destructive, meaning
that these large bodies are able to rapidly deplete the population of small particles
that remain in the disk.

\begin{figure}
\includegraphics[width=3.1in]{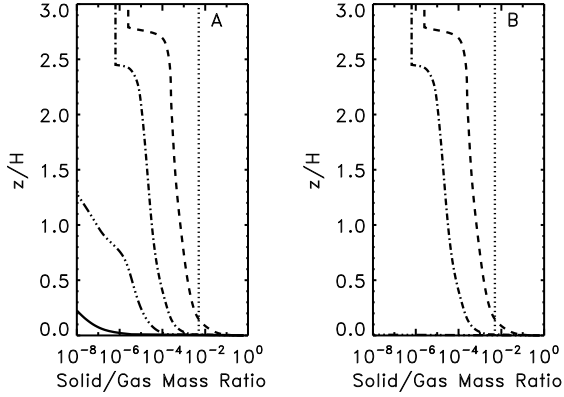}
\caption{Vertical distributions of solids at 1 AU in a layered protoplanetary
disk for a laminar dead zone (A) and a dead zone which gets stirred up from
the active region above (B).  The lines represent the same times described in 
Figure 1. The 10$^{5}$ and 10$^{6}$ year marks for case B are absent as all
the mass of solids has been incorporated into large bodies at the midplane at
these times.}
\end{figure}

\section{Discussion}

The results presented here show that in layered protoplanetary disks, the active
regions above the dead zone become depleted of dust on short 
timescales, particularly when compared to the cases of a uniformly laminar or
turbulent disk. 
While radial transport has been neglected, the timescale for dust
to be delivered to the dead zone from above is short.  Thus, it is unlikely that
radial transport would be sufficient to replenish dust in the active layers to
levels far above those seen here.  Even turbulence generated by a particle rich
layer at the midplane would be ineffective as the vertical scale of such turbulence is expected to be
much less than that considered here \citep{weidenschilling97}.
This depletion of solids at high altitudes will have important consequences for 
modeling and interpreting observations of protoplanetary disks.

Dust at high altitudes (3-5 scale heights)  absorbs and
scatters radiation from the central star and determines how radiation is
processed by the disk \citep{chiang01, dalessio06}.  If dust is
absent from this region, or even depleted, it will drastically affect this
processing.  In particular, \citet{dalessio06} found that 
while dust depletions in the upper few scale heights would not alter the surface
temperatures of the disk, midplane temperatures would decrease because less 
radiation would be processed and directed to the disk interior.  This would
be particularly important in dead zones which do not generate heat from viscous 
dissipation.

\begin{figure}
\includegraphics[width=2.8in,angle=90]{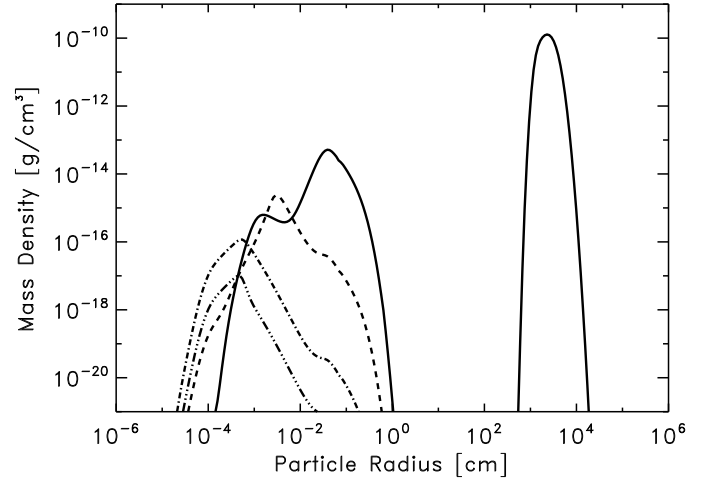}
\caption{Plotted are the mass density of particles in a layered disk when the
MRI-active region grows massive enough to "stir up" turbulence in the dead zone.
The lines represent the distribution of particles at the midplane (solid line) at
one scale-height ($H$) (dashed), 2$H$ (dash-dotted), and 3$H$ (dash-dot-dot-dotted).}
\end{figure}

The radial extent of the dead zone would vary from disk to disk, and depends
on such factors as the strength of the magnetic field, the surface density
distribution of the disk, the size of the dust particles, and their composition
\citep{sano00}.  However, the general expectation is that beyond the point where the
dead zone terminates, the disk will be MRI-active at all heights, and therefore
uniformly turbulent.  Thus, this would create a situation where dust is depleted
in the inner disk, but present, and at high altitudes, in the outer parts of
the disk.  Such structures have been inferred for disks around TW Hydra, GM
Auriga, and CoKu Tau/4 \citep{calvet05b,dalessio05}.  These disks appear to be
significantly depleted in dust in the inner disk, but are optically thick at
larger radii.  The transition between these two regions is thought to be very
sharp, with the the inner edge, or wall, of the optically thick region being directly
illuminated by the central star.  The clearing of the inner parts of the disk
was originally attributed to a planet opening a gap, but it is thought that
the high residual mass of the disk would lead to rapid inward migration due
to disk torques \citep{calvet05b}.  

Based on the results presented here, it is possible that in some disks, the inferred wall 
represents the termination of the dead zone in the disk.  Inside of the wall, 
most of the solids have settled to the midplane and have been incorporated into
planetesimals, whereas outside of the wall, the growth of large bodies has been
inhibited by the presence of turbulence, as in case B of Figure 1.  Not only
would this explain why the outer regions are optically thick, but it would also
explain why dust is present at such high altitudes outside of the wall, but
nearly absent inside of it.  The minor amount of dust inside of the wall in the
disks around TW Hydra and GM Auriga may be tied to the fact that these disks
are still accreting onto the central star.  This implies that non-negligible
mass transport is taking place through the disk \citep{calvet05b}, and thus, 
small grains may be carried inwards from the optically thick outer disk by
the net advective flows associated with disk evolution.  CoKu Tau/4, on the
other hand, is not accreting significant amounts of mass, and therefore would
be unable to resupply the inner region with fine dust at a significant rate.
For those disks in which no obvious inner gap appears, it is possible that either
the dead zone spans a negligible region of the disk, or that the dead zone is
being stirred up by other effects.  For example, gravitationally unstable disks
have been shown to be capable of generating turbulence locally or to produce 
large-scale transport of fine dust on short timescales \citep{boss04,boley06}.  

These results also offer a potential way of interpretting the meteoritic record
of how solids grew in our own solar nebula.  Recent results suggest that many
iron meteorites are just as old as the oldest known objects in our 
solar system, the Calcium-Aluminum-rich Inclusions (CAIs) \citep{kleine05}.  
Previously, it was thought that chondritic meteorite parent bodies represented 
the oldest accreted planetesimals in our solar system.  With chondrules, the dominant 
components of chondritic meteorites, being typically 1-3 million years younger
than CAIs, nebular turbulence had been invoked as a way of delaying 
the formation of large bodies until this time period.  It has also been argued
that other properties of chondritic meteorites, such as the narrow size range of
chondrules in a given meteorite, are easier to understand if they formed in a 
turbulent nebula \citep{cuzzi01,cuzzi05}.  

The layered disk model presented here offers a possible way to allow for the 
early formation of large bodies and the subsequent formation of chondritic 
meteorite parent bodies in a turbulent environment.  Large bodies could form
during the initial stages of disk evolution when the MRI operated only in
the extreme upper layers of the disk.  The midplane region of the disk would
be laminar and allow for the rapid formation of kilometer-sized bodies on
relatively short timescales.  These bodies could grow further through gravitational
effects, and would likely contain enough $^{26}$Al and $^{60}$Fe that they 
would differentiate and develop iron cores.  As the MRI-active region in the 
upper layers of the disk grew due to dust depletion and surpassed the critical size
as given by \citet{fleming03}, the entire interior of the disk would become
turbulent.   While the
case considered here suggested that the dust that remained in the disk when the
dead zone became turbulent would be rapidly accreted by the larger bodies, this
treatment neglected the effects of chondrule formation events, or the possible
formation of chondrite parent bodies directly from dust by turbulent concentration
\citep[i.e.][]{cuzzi01}.  Additionally, if turbulence is present in the disk
midplane, mass and angular momentum transport could occur through the dead zone, 
allowing dust to be resupplied to the inner disk from larger heliocentric
distances.  This would provide a turbulent environment filled with matrix
and chondrule precursors in which chondritic meteorite parent bodies could
form.

In conclusion, the upper layers of turbulently layered protoplanetary disks are depleted in
dust more rapidly than either uniformly laminar or uniformly turbulent disks.  This
is because the turbulence in the upper layers of a disk leads to the rapid production
of large bodies which are then stored in the laminar midplane region of the disk.
Even if the midplane region later becomes turbulent because it is stirred up from
above, large bodies may grow from the ``seeding'' that occurs
during the initial laminar phase.  While the particular results of the model
are parameter dependent, there are aspects of protoplanetary disk observations
and the meteorite record of our own solar system that can be understood in
the context of a layered disk.  Future studies should investigate these effects
in detail.

\acknowledgments
FJC thanks John Chambers for helpful discussions.  An anonymous reviewer also
made very insightful suggestions that led to a much improved manuscript and
investigation.  This work was supported by the Carnegie Institution of Washington.

%\bibliographystyle{apj}
%\bibliography{neb_apj}

\end{document}